\documentclass[%
reprint,
superscriptaddress,
nofootinbib,
amsmath,amssymb,
aps,
prb,
]{revtex4-2}
\usepackage{booktabs}
\usepackage{graphicx}% Include figure files
\usepackage{dcolumn}% Align table columns on decimal point
\usepackage{bm}% bold math
\usepackage{chemformula}
\usepackage{hyperref}% add hypertext capabilities
\usepackage{amsfonts,amssymb}
\usepackage{makecell}
\usepackage{hyperref}% add hypertext capabilities
\hypersetup{hidelinks,
	colorlinks=true,
	allcolors=blue,
	pdfstartview=Fit,
	breaklinks=true}

\usepackage{tabularx}% Table width
\usepackage{amsmath}% Higher math
\usepackage{float}
\begin{document}
	\title{Absence of Acoustic Phonon Anomaly in a Kagome Metal with Short-ranged Structural Modulation}
	\author{Weiliang~Yao}
	\affiliation{Department of Physics and Astronomy, Rice University, Houston, Texas 77005, USA}
	\author{Supeng~Liu}
	\affiliation{Department of Advanced Materials Science, The University of Tokyo, Kashiwa, Chiba 277-8561, Japan}
	\author{Zifan~Xu}
	\affiliation{Department of Advanced Materials Science, The University of Tokyo, Kashiwa, Chiba 277-8561, Japan}
	\author{Daisuke~Ishikawa}
	\affiliation{Materials Dynamics Laboratory, RIKEN SPring-8 Center, RIKEN, Sayo, Hyogo 679-5148, Japan}
	\affiliation{Precision Spectroscopy Division, Japan Synchrotron Radiation Research Institute, Kouto 1-1-1, Sayo, Hyogo 679-5198, Japan}
	\author{Zehao~Wang}
	\affiliation{Department of Physics and Astronomy, Rice University, Houston, Texas 77005, USA}
	\author{Bin~Gao}
	\affiliation{Department of Physics and Astronomy, Rice University, Houston, Texas 77005, USA}
	\author{Sijie~Xu}
	\affiliation{Department of Physics and Astronomy, Rice University, Houston, Texas 77005, USA}
	\author{Feng~Ye}
	\affiliation{Neutron Scattering Division, Oak Ridge National Laboratory, Oak Ridge, Tennessee 37831, USA}
	\author{Kenichiro~Hashimoto}
	\affiliation{Department of Advanced Materials Science, The University of Tokyo, Kashiwa, Chiba 277-8561, Japan}
	\author{Takasada~Shibauchi}
	\affiliation{Department of Advanced Materials Science, The University of Tokyo, Kashiwa, Chiba 277-8561, Japan}
	\author{Alfred~Q.~R.~Baron}
	\affiliation{Materials Dynamics Laboratory, RIKEN SPring-8 Center, RIKEN, Sayo, Hyogo 679-5148, Japan}
	\author{Pengcheng~Dai}
	\affiliation{Department of Physics and Astronomy, Rice University, Houston, Texas 77005, USA}
	\affiliation{Smalley-Curl Institute, Rice University, Houston, Texas 77005, USA}
	
	\date{\today}
	
	\begin{abstract}
		Kagome lattice $A$V$_3$Sb$_5$ ($A$ = K, Rb, and Cs) superconductors without magnetism from vanadium $d$-electrons are intriguing because they have a novel charge density wave (CDW) order around 90 K and display superconductivity at $\sim$3 K
		that competes with the CDW order. Recently, CsCr$_3$Sb$_5$, isostructural to $A$V$_3$Sb$_5$, was found to have concurrent structural and magnetic phase transition at $T^{\ast}\approx$ 55 K that can be suppressed by pressure to induce superconductivity [Liu \textit{et al.}, \href{https://doi.org/10.1038/s41586-024-07761-x}{Nature \textbf{632}, 1032 (2024)}]. Here, we use elastic and inelastic X-ray scattering to study the microscopic origin of the structural transition in CsCr$_3$Sb$_5$. Although our elastic measurements confirm the 4$\times$1$\times$1 superlattice order below $T^{\ast}$, its underlying correlation is rather short-ranged. Moreover, our inelastic measurements at the superlattice wavevectors around (3, 0, 0) find no evidence of a significant acoustic phonon anomaly below $T^{\ast}$, similar to the case of $A$V$_3$Sb$_5$. The absence of acoustic phonon anomaly indicates a weak electron-phonon coupling in CsCr$_3$Sb$_5$, suggesting that the structural transition is likely associated with an unconventional CDW order.
	\end{abstract}
	
	\maketitle
	Understanding the intertwined orders of charge-spin-lattice degrees of freedom in quantum materials forms the basis to unveil the microscopic origin of their exotic electronic properties \cite{KeimerNP2017,KeimerNature2015,FradkinRMP2015,TranquadaPhysicaB2015,DaiRMP2015}. For example, in copper-oxide superconductors, the interplay between charge density wave (CDW) and magnetism may mediate electron pairing for superconductivity \cite{KeimerNature2015,FradkinRMP2015,TranquadaPhysicaB2015}, clearly different from the electron-phonon coupling (EPC) induced superconductivity in conventional Bardeen-Cooper-Schrieffer (BCS) superconductors \cite{BardeenPR1957}. 
	
	Recently, kagome lattice metals have become an appealing platform to study the intertwined charge-spin-lattice order because they display flat electronic bands, Dirac points, Van Hove singularities, and can have interplay between CDW, magnetic order, and superconductivity \cite{YinNature2022,WangNatRevPhys2023}. For weakly electron correlated kagome lattice superconductors $A$V$_3$Sb$_5$ ($A$ = K, Rb, and Cs), CDW order coexists and competes with superconductivity but without magnetism from vanadium $d$-electrons \cite{WilsonNRM2024}. Electron correlated kagome magnet FeGe, on the other hand, has coupled CDW and magnetic order but no superconductivity \cite{TengNature2022}. 
	
\begin{figure}[t!]
		\includegraphics[width=9cm]{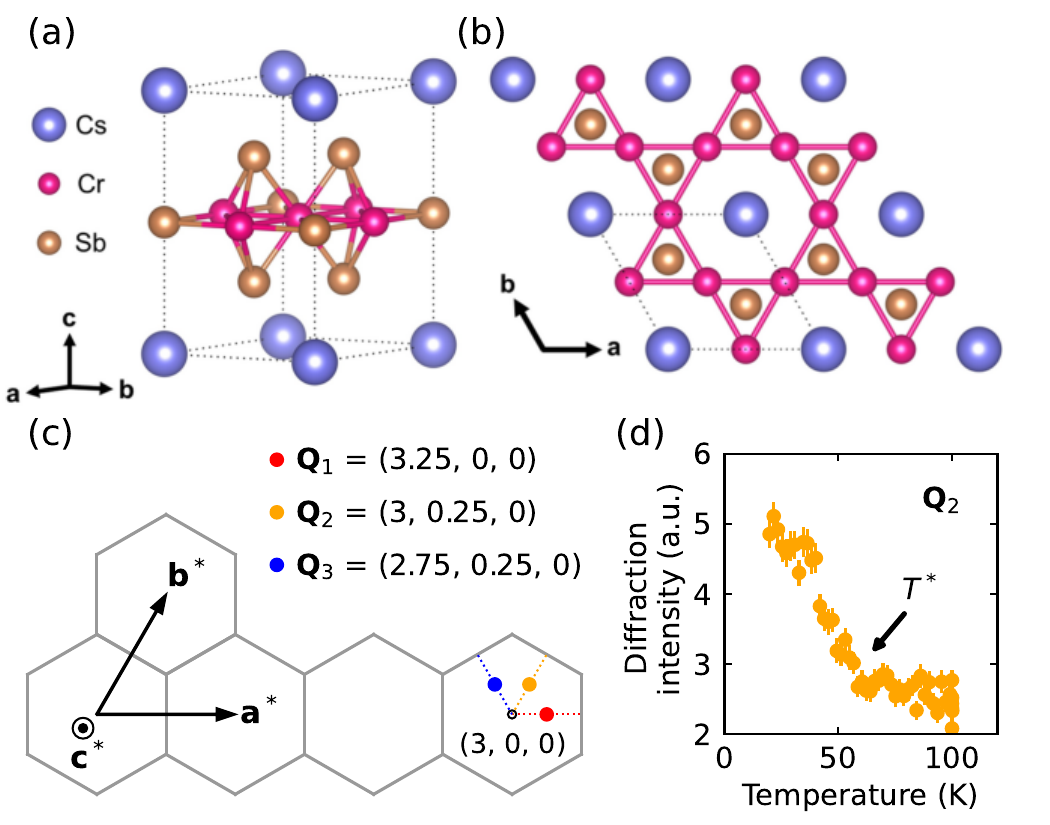}
		\caption{(a) and (b) Crystal structure of CsCr$_3$Sb$_5$. The unit cell is marked by dashed lines. (c) Sketch of reciprocal space. The hexagons represent the Brillouin zone boundaries. The momentum transfer positions $\mathbf{Q}_1$, $\mathbf{Q}_2$, and $\mathbf{Q}_3$ are marked with red, orange, and blue dots, respectively. (d) Temperature dependence of the diffraction intensity at $\textbf{Q}_2$.}
		\label{fig1}
	\end{figure}

	The newly synthesized electron correlated kagome metal CsCr$_3$Sb$_5$ was found to display a concurrent structural and magnetic phase transition below $T^{\ast} \approx$ 55 K \cite{LiuNature2024}. Figure \ref{fig1}(a) and (b) show the crystal structure of CsCr$_3$Sb$_5$. As a sister compound of $A$V$_3$Sb$_5$ \cite{WilsonNRM2024}, it crystallizes into the space group $P6/mmm$ (no. 191). Cr atoms form a kagome sublattice, which is coordinated by the adjacent Sb atoms and is further intercalated by the Cs triangular sublattices. Although structurally similar, the physical properties of CsCr$_3$Sb$_5$ show noteworthy distinctions from $A$V$_3$Sb$_5$. It has a CDW-like structural modulation occurring at $T^{\ast}$, which can be described by the wave vector $(0.25, 0, 0)$ (in hexagonal basis), and its equivalences $(0, 0.25, 0)$ and (-0.25, 0.25, 0) [Fig. \ref{fig1}(c)] due to the C$_6$ symmetry of the lattice \cite{LiuNature2024}. This structural modulation is found to be accompanied by an antiferromagnetic spin density wave (SDW) order \cite{LiuNature2024}. Moreover, under moderately high pressures of 4-8 GPa, these density waves are suppressed, and unconventional superconductivity with a maximum transition temperature of $\sim$6.4 K emerges \cite{LiuNature2024}. Since magnetic order in CsCr$_3$Sb$_5$ is most likely arising from the localized magnetic moments of Cr $d$-electrons due to electron correlations \cite{LiuNature2024}, the system is clearly different from the nonmagnetic $A$V$_3$Sb$_5$ with weak electron correlations \cite{WilsonNRM2024}.
	
	Despite CsCr$_3$Sb$_5$ exhibits remarkably interesting properties, the origin of the transition at $T^*$ is far from clear \cite{LiuNature2024}. If the structural lattice distortion below $T^*$ is associated with a conventional CDW order \cite{LiuNature2024}, one would expect to observe EPC through modifications of acoustic phonon modes across the underlying structural transition temperature $T^*$ \cite{ZhuPNAS2015,ZhuAPX2017}. Inelastic X-ray scattering (IXS) can directly measure the dynamic structure factor of phonons \cite{Baron2020}, which is therefore an ideal experimental probe to study the origin of the structural distortion. For BCS superconductors, studying acoustic phonons across the superconducting temperature is useful, as superconductivity-induced phonon softening and phonon resonance via EPC can sometimes be observed \cite{AxePRB1965,KawanoPRL1996}.
	
	In this work, we present an IXS study on CsCr$_3$Sb$_5$ single crystals. On cooling across $T^{\ast}$, our elastic X-ray scattering measurements find broad 4$\times$1$\times$1 superlattice peaks, suggesting short-ranged nature of the structural modulation. Using IXS, we mapped out low-energy phonon spectra below and above $T^{\ast}$. The acoustic phonons around the superlattice wave vector are basically unchanged across $T^{\ast}$. These results are similar to the behaviors of acoustic phonons in CDW ordered $A$V$_3$Sb$_5$ \cite{LiPRX2021,XiePRB2022,SubiresNC2023} and FeGe \cite{MiaoNC2023,TengNP2023,TengPRL2024}, thus indicating that the EPC is quite weak and not the driving force for the CDW-like transition at $T^{\ast}$ in CsCr$_3$Sb$_5$.
	
	CsCr$_3$Sb$_5$ single crystals were grown with a flux method as reported earlier \cite{LiuNature2024,GuoArxiv2024}. A piece of CsCr$_3$Sb$_5$ single crystal ($\sim$$0.5\times0.3\times0.003$ mm$^3$) was mounted on the tip of a copper sample holder in a transmission geometry \cite{SM}. The observed overall sample mosaic was about 0.3 degrees \cite{SM}. The IXS experiment was performed using the meV-resolution spectrometer at BL43LXU \cite{Baron2010} of SPring-8, in Hyogo prefecture, Japan. This spectrometer provides world-leading flux with excellent energy resolution \cite{Baron2010}. In our experiment, the Si(12,12,12) setup was used with a 4 $\times$ 7 array of analyzer crystals, allowing 28 momentum transfer positions to be measured simultaneously.  The energy resolution was between 1.15 and 1.35 meV for most of the analyzers. The momentum transfer position is defined as $\mathbf{Q}$ = $H\mathbf{a}^{\ast} +K\mathbf{b}^{\ast}+L\mathbf{c}^{\ast}$ [Fig. \ref{fig1}(c)], where $a^\ast=b^\ast=\frac{4\pi}{\sqrt3 a}$, $c^\ast=\frac{2\pi}{c}$, and $a$, $b$, $c$ are the lattice constants. The momentum transfer resolution was set using 30 $\times$ 30 mm$^2$ slits at 9 m from the sample to be $\sim$(0.035, 0.035, 0.015) reciprocal lattice units in most measurements. Typical data collection time is $\sim$90 minutes/spectrum, despite the low signal rate due to the very thin sample.
	
	\begin{figure}[t!]
		\includegraphics[width=8.8cm]{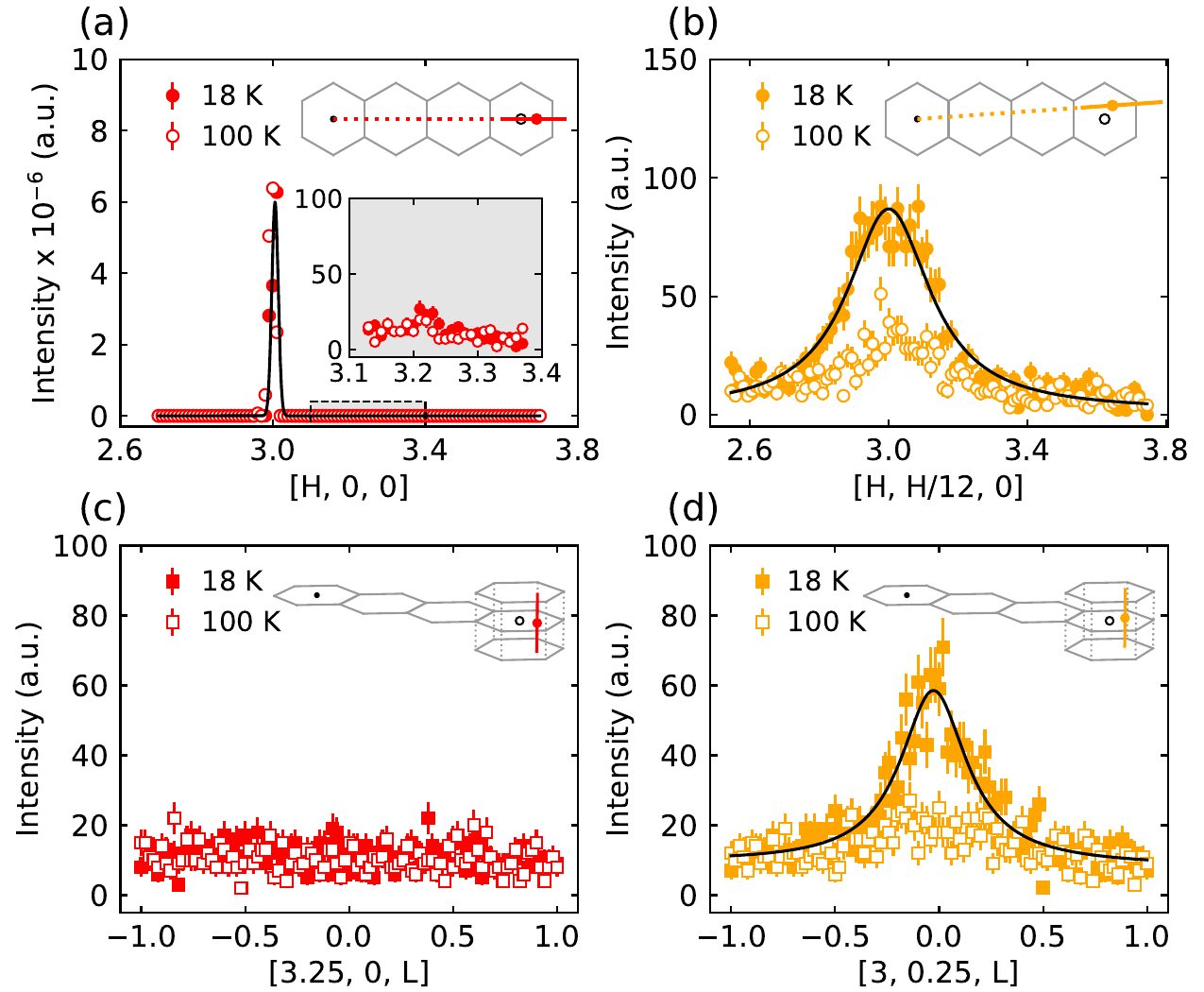}
		\caption{(a) and (b) $\theta$-2$\theta$ diffraction scans of 18 K and 100 K along trajectories going through $\mathbf{Q}_1$ and $\mathbf{Q}_2$, respectively. The middle inset is a zoom-in view around $\mathbf{Q}_1$ (dashed box). (c) and (d) $L$-scans of 18 K and 100 K along trajectories going through $\mathbf{Q}_1$ and $\mathbf{Q}_2$, respectively. These scans were done by setting the energy transfer to zero. The upper right insets in each panel show their trajectories in the reciprocal space. Solid curves in (a), (b), and (d) are fits to the data of 18 K as described in the text.}
		\label{fig2}
	\end{figure}
	
	Fig. \ref{fig1}(d) shows the temperature dependence of diffraction intensity at the superlattice peak $\mathbf{Q}_2 = (3, 0.25, 0)$. The intensity abruptly increases below $\sim$55 K, which is consistent with the phase transition observed in previous resistivity, magnetization, and specific heat measurements \cite{LiuNature2024}. To study the structural modulation, we performed $\theta$-2$\theta$ diffraction scans along the [$H$, 0, 0] and [$H$, $H$/12, 0] directions, which go through two momentum transfer positions $\mathbf{Q}_1 = (3.25, 0, 0)$ and $\mathbf{Q}_2$. In Figure \ref{fig2}(a), we see the strong and sharp fundamental Bragg peak $(3, 0, 0)$, which is temperature-independent. However, there is no observable peak at $\mathbf{Q}_1$, as can be more clearly seen from the inset of Fig. \ref{fig2}(a). On the other hand, the scan through $\mathbf{Q}_2$ shows a peak at 18 K, which is $\sim$4 orders of magnitude weaker than the fundamental one, comparable to the case of $A$V$_3$Sb$_5$ \cite{LiPRX2021}. Some intensities remain but are much weakened at 100 K, probably due to leaked inelastic signal or residual critical scattering \cite{BirgeneauPRB1971,ShapiroPRB1972,MonctonPRB1977}, as the transition at $T^{\ast}$ is second-order or weakly first-order \cite{LiuNature2024}. We also confirm that there is a peak at $\mathbf{Q}_3 = (2.75, 0.25, 0)$ below $T^{\ast}$ \cite{SM}. Such observation is consistent with the diffraction feature that four out of six superlattice peaks around $(3, 0, 0)$ are evident, which was previously found based on an in-house X-ray diffraction experiment \cite{LiuNature2024}. If the observed structural modulation is induced by a CDW transition, we should expect the most pronounced acoustic phonon anomalies at the momentum transfer positions associated with the structural modulation regardless of the strength of EPC \cite{MonctonPRL1975}.
	
	\begin{figure}[t!]
		\includegraphics[width=8.5cm]{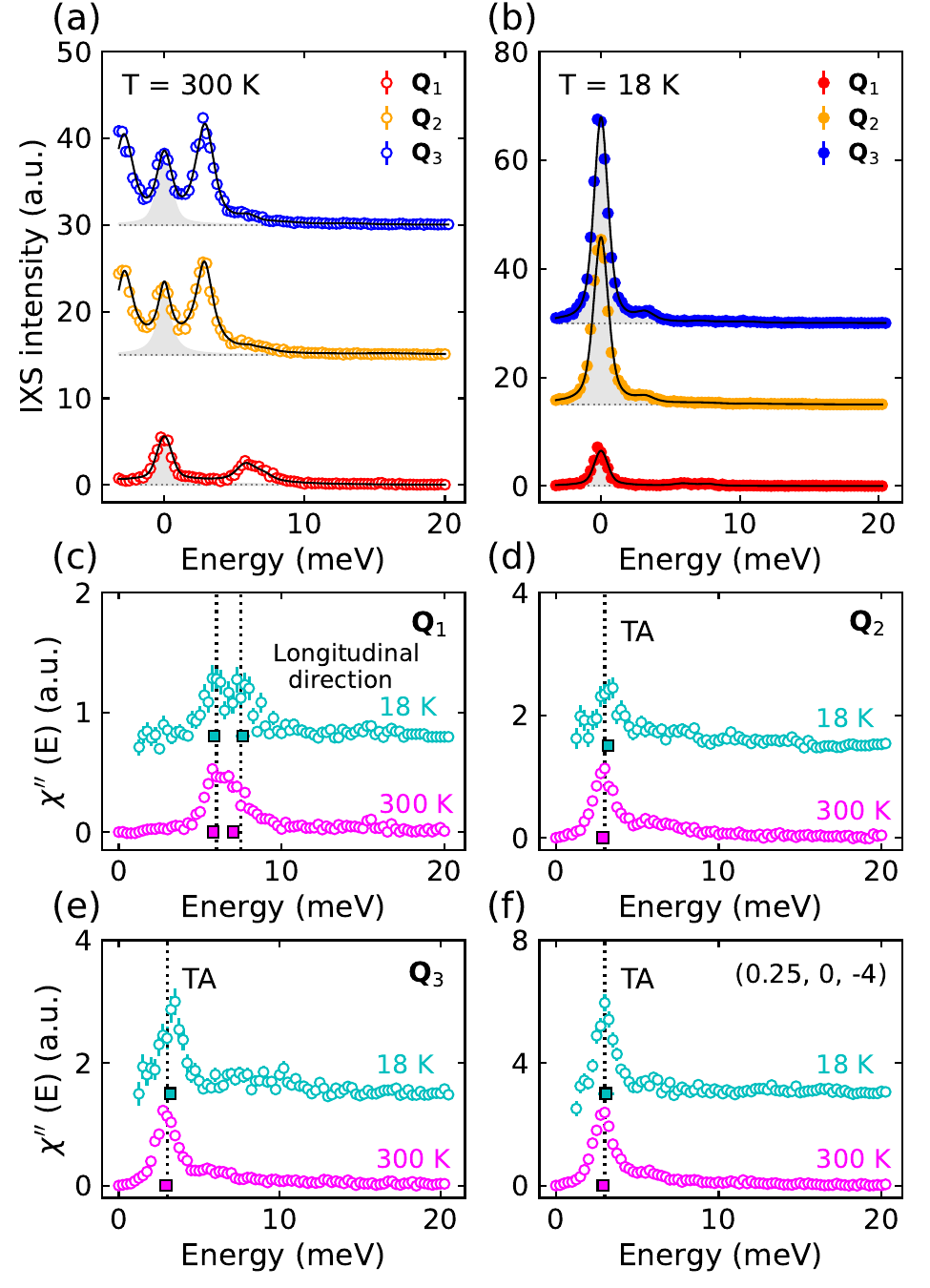}
		\caption{(a) and (b) Energy dependence of measured IXS intensities of $\mathbf{Q}_1$, $\mathbf{Q}_2$, and $\mathbf{Q}_3$ at 18 K and 300 K, respectively. Data are evenly offset for clarity. Zero intensity for the corresponding data is indicated by dashed horizontal lines. The solid curves are the fits to the data as described in the text. The elastic scattering contributions are indicated by the shaded area. (c)-(f) Energy dependence of the imaginary part of the dynamic susceptibilities of 300 K and 18 K at $\mathbf{Q}_1$, $\mathbf{Q}_2$, $\mathbf{Q}_3$, and (0.25, 0, -4), respectively. Vertical dashed lines in (d)-(f) indicate the TA mode energies, while in (c) two modes are clearly visible (see text).  Filled squares show the fitted phonon energies as described in \cite{SM}. The error bars are close to or smaller than the size of the data symbol.}
		\label{fig3}
	\end{figure}
	
	We further check the diffraction profiles along the $L$-direction. Fig. \ref{fig2}(c) presents the data with the momentum transfer positions going through $\mathbf{Q}_1$, which show the absence of a peak both at 18 K and 100 K. Along the $[3, 0.25, L]$ direction, there is a peak centering at $\mathbf{Q}_2$ [Fig. \ref{fig2}(d)], consistent with the scan within the kagome plane. The $\mathbf{Q}_2$ peak at 18 K are broad along the [$H$, $H$/12, 0] and [3, 0.25, $L$] directions, indicating short-ranged correlations in both directions. We fit these peaks with a Lorentzian profile [Fig. \ref{fig2}(b) and (d)] and estimate the correlation lengths from the peak width. The resultant correlation lengths along and perpendicular to the kagome plane are $\xi_{\parallel} = 33(2)$ \AA \,($\approx 6.0a$) and $\xi_{\perp} = 48(4)$ \AA \,($\approx 5.2c$), respectively. These correlation lengths are much smaller than the one based on the structural Bragg peak (3, 0, 0) [Fig. \ref{fig2}(a)], which is 5.7(3) $\times$ 10$^2$ \AA \,($\approx 100a$) based on the fit with a Gaussian profile and roughly represents the instrumental momentum resolution.
	
	Next, we turn to the IXS-measured phonon spectra. We mainly took the data around the Brillouin zone centering at (3, 0, 0) [Fig. \ref{fig1}(c)], which has a strong X-ray scattering structure factor \cite{LiuNature2024}. Figure \ref{fig3}(a) and (b) show the IXS spectra at $\mathbf{Q}_1$, $\mathbf{Q}_2$, and $\mathbf{Q}_3$ at 300 K and 18 K, respectively. While the elastic scattering ($E$ = 0 meV) intensities at these three positions are comparable at 300 K, the intensities at $\mathbf{Q}_2$ and $\mathbf{Q}_3$ significantly increase at 18 K, and the scattering intensity at $\mathbf{Q}_1$ remains largely the same. Apart from the peaks located at $E = 0$ meV, we can clearly discern one major peak between 0 and 10 meV for all three positions at 300 K [Fig. \ref{fig3}(a)], whose intensities are greatly reduced at 18 K [Fig. \ref{fig3}(b)]. We ascribe these peaks to acoustic phonon modes. In an IXS experiment, the measured intensity is proportional to the dynamic structure factor $S (\mathbf{Q}, E)$ \cite{Baron2020}. Following the standard procedure, we fit the measured $S (\mathbf{Q}, E)$ with a sum of the signals from different phonon modes multiplied by the Bose factor and convoluted with the instrumental energy resolution function \cite{MiaoPRX2018,DashwoodPRB2019,LiPRX2021,CaoNC2023,KorshunovNC2023,SongPRB2023}. The fitting results are shown with the solid curves in Fig. \ref{fig3}(a) and (b). More details of the fit can be found in \cite{SM}.
	
\begin{figure}[t!]
		\includegraphics[width=8.5cm]{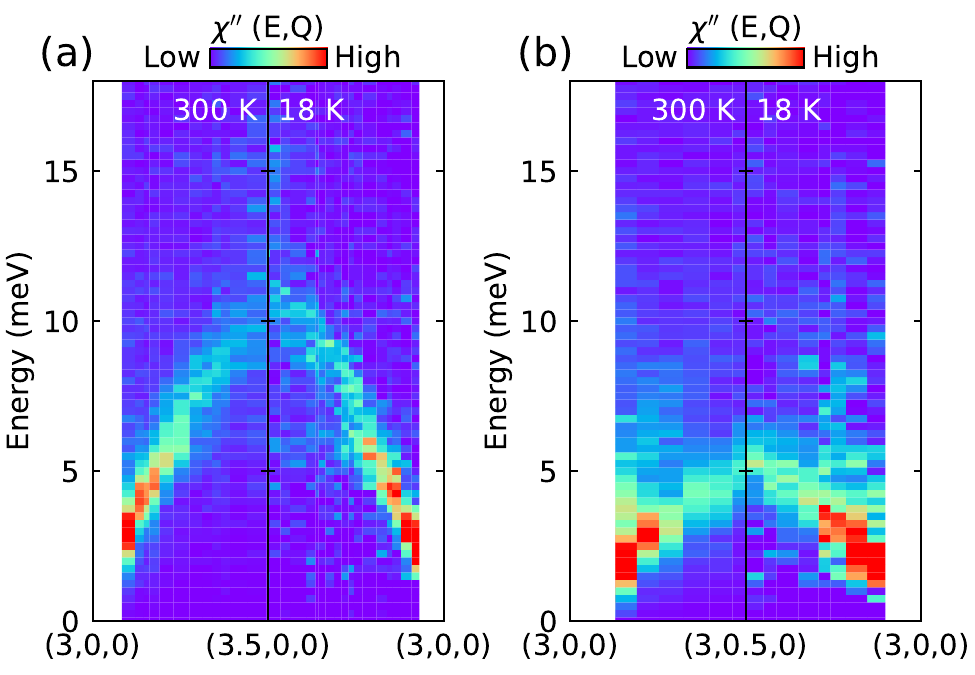}
		\caption{(a) Color maps of the imaginary part of the dynamic susceptibilities at 300 K (left) and 18 K (right) from (3, 0, 0) to (3.5, 0, 0). (b) Color maps of the imaginary part of the dynamic susceptibilities at 300 K (left) and 18 K (right) from (3, 0, 0) to (3, 0.5, 0). The intensities are shown in a linear scale.}
		\label{fig4}
	\end{figure}

	After removing the elastic contribution [shaded parts in Fig. \ref{fig3}(a) and (b)], we get the measured imaginary part of the dynamic susceptibility $\chi^{\prime\prime} (\mathbf{Q},E)$ [Fig. \ref{fig3}(c)-(f)],  which reflects the intrinsic properties of the phonon. At 18 K, $\chi^{\prime\prime} (\mathbf{Q_1},E)$, in a purely longitudinal geometry, shows two resolvable peaks around 6 meV and 7.5 meV [Fig. \ref{fig3}(c)], which slightly merge at 300 K while the combined peak width is always significantly larger than the energy resolution. This peak doubling is probably due to the presence of a low-lying optical phonon mode, as is supported by recent calculations \cite{XuArxiv2023}, that anti-crosses with the longitudinal acoustic (LA) phonon mode. Even so, the LA phonon dispersion is clear (see discussion below and Figure \ref{fig4}).  Fig. \ref{fig3}(d) and (e) show $\chi^{\prime\prime} (\mathbf{Q_2},E)$ and $\chi^{\prime\prime} (\mathbf{Q_3},E)$, respectively, which show the transverse acoustic (TA) phonon mode around 3 meV with almost resolution limited peak widths. For both the LA and TA modes, we notice that although their energies shift a bit with temperature, the changes are all within the error bar [Fig. \ref{fig3}(c)-(e)]. Moreover, we find that the data at 18 K almost overlap with those at 300 K, which therefore rules out a significant acoustic phonon anomaly across $T^{\ast}$. This is confirmed by the measurement at (0.25, 0, -4) [Fig. \ref{fig3}(f)], which is an equivalent position with $\mathbf{Q_1}$, but has a large $c$-axis component so sees $c$-axis polarized motion - this mode also shows essentially no change with temperature.
	
	With more measurements at additional momentum transfer positions, we construct a color map of the imaginary part of the dynamic susceptibility from (3, 0, 0) to (3.5, 0, 0) and from (3, 0, 0) to (3, 0.5, 0), as shown in Fig. \ref{fig4}. These two trajectories pass through $\mathbf{Q}_1$ and $\mathbf{Q}_2$, respectively [Fig. \ref{fig1}(c)]. The whole energy bandwidths are about 10 meV for the LA phonon and about 5 meV for the TA phonon. These phonon bandwidths are similar to those of CsV$_3$Sb$_5$ \cite{XiePRB2022,SubiresNC2023} but slightly smaller than RbV$_3$Sb$_5$ \cite{LiPRX2021}. We note that these non-collapsed acoustic phonons at ambient pressure are in line with a recent first-principle calculation, which is based on an antiferromagnetic ground state \cite{XuArxiv2023}. The similar phonon dispersions at 18 K and 300 K further corroborate the notion that there is no significant anomaly below $T^{\ast}$ for both LA and TA phonons. From the long-wavelength limit of the LA phonon, we can estimate the longitudinal sound velocity to be 3.2(1) $\times$ 10$^3$ m$\cdot$s$^{-1}$ along the kagome plane, which is close to the in-plane longitudinal sound velocity in CsV$_3$Sb$_5$ [3.6(2) $\times$ 10$^3$ m$\cdot$s$^{-1}$] \cite{XiePRB2022} but significantly smaller than the one in FeGe [5.2(2) $\times$ 10$^3$ m$\cdot$s$^{-1}$] \cite{TengNP2023,TengPRL2024}. These sound velocities along with the densities show that CsCr$_3$Sb$_5$ has the smallest Young’s modulus among the three kagome metals, and therefore is the softest and most flexible.
	
	Having a CDW-like transition but without an acoustic phonon anomaly is quite unusual. For CsCr$_3$Sb$_5$, the correlation length of the CDW-like structural modulation is within 10 unit cells, which looks to be insufficient to impact the propagation of the lattice vibration. However, acoustic phonon anomaly across the CDW transition was actually not uncommon for materials with comparable CDW correlation lengths \cite{LeTaconNP2014,MiaoPRX2018,Ziebeck1977}. For example, in the nearly commensurate CDW state of 1$T$-TaS$_2$, where the structural modulation is spatially isolated Star-of-David clusters \cite{WilsonAdvPhys1975,RossnagelJPCM2011}, the acoustic phonon still shows an appreciable anomaly across CDW transition \cite{Ziebeck1977}. Since the acoustic phonon anomaly is generally induced through EPC, the absence of such anomaly in CsCr$_3$Sb$_5$ therefore excludes strong EPC and indicates its CDW-like transition is unconventional \cite{ZhuPNAS2015,ZhuAPX2017,LiPRX2021}.
	
	Interestingly, the absence of significant acoustic phonon anomaly was also found in nonmagnetic kagome metals $A$V$_3$Sb$_5$ and magnetic FeGe \cite{LiPRX2021,XiePRB2022,TengNP2023,SubiresNC2023,TengPRL2024}, where the unconventional CDWs are competing with superconductivity and antiferromagnetism, respectively \cite{OrtizPRM2019,OrtizPRL2020,ChenPRL2021,TengNature2022}. However, the superlattice of CsCr$_3$Sb$_5$ is 4$\times$1$\times$1 type and shows short-ranged correlation \cite{LiuNature2024}, which differs from the 2$\times$2$\times$2 superlattice in both $A$V$_3$Sb$_5$ and FeGe \cite{OrtizPRL2020,LiPRX2021,TengNature2022,SubiresNC2023,MiaoNC2023}. We note that recent angle-resolved photoemission spectroscopy studies on CsCr$_3$Sb$_5$ have revealed a Fermi wave vector $k_{\rm{F}} \approx 0.2 a^{\ast}$ \cite{LiArxiv2024,GuoArxiv2024,PengArxiv2024}. For a Fermi surface nesting scenario, one expects an acoustic phonon anomaly at 2$k_{\rm{F}}$ with an accompanied structural modulation \cite{ZhuPNAS2015,ZhuAPX2017}, which is, however, not observed in our experiment. Therefore, we can rule out Fermi surface nesting as the major driving force for the CDW-like structural modulation in CsCr$_3$Sb$_5$ \cite{PengArxiv2024}.
	
	Although the current study focuses on acoustic phonons with dominant spectral weight, subtle phonon anomalies might still be present in CsCr$_3$Sb$_5$, particularly in the high-energy optical modes or other unexamined regions of the reciprocal space. Indeed, interesting optical phonon anomalies were observed in the CDW state of $A$V$_3$Sb$_5$ \cite{LiPRX2021,XiePRB2022} and FeGe \cite{TengPRL2024}, even in the absence of a significant anomaly in the acoustic modes. Nevertheless, the short-ranged structural modulation and the absence of acoustic phonon anomaly across $T^{\ast}$ have two implications on the physical properties of CsCr$_3$Sb$_5$. First, the phase transition at $T^{\ast}$ is evident in resistivity and magnetization \cite{LiuNature2024}, but has minor effects on the lattice dynamics. Instead, the $T^{\ast}$-transition may exert a more profound influence on magnetic order and spin dynamics, as seen in the CDW order of FeGe \cite{TengPRL2024}. Second, the CDW-like order in CsCr$_3$Sb$_5$ is suppressed by applying pressure and finally gives way to the superconductivity \cite{LiuNature2024}. As no significant acoustic phonon anomaly is observed at ambient pressure and the phase boundary continuously ends in the superconductivity dome, it would be reasonable to assume that EPC is also weak in the high-pressure superconducting state. This would imply that superconductivity in CsCr$_3$Sb$_5$ is not mediated by EPC but may arise from magnetism as in the case of cuprates and iron pnictides \cite{KeimerNature2015,FradkinRMP2015,TranquadaPhysicaB2015,DaiRMP2015}. This aspect is likely different from $A$V$_3$Sb$_5$ without magnetism \cite{WilsonNRM2024}, where $s$-wave superconductivity may be mediated by EPC under the conventional BCS mechanism \cite{MuCPL2021,ZhongNC2023,XieNC2024} or by bond-order fluctuations \cite{RoppongiNC2023,TazaiSA2022}. To address this point, it is necessary to determine the pairing symmetry in CsCr$_3$Sb$_5$ under high pressure.
	
	In summary, by performing an IXS experiment on the single crystal of kagome metal CsCr$_3$Sb$_5$, we have successfully mapped out its acoustic phonon spectra along high-symmetric directions in the hexagonal Brillouin zone. We find that the CDW-like structural modulation below $T^{\ast}$ is short-ranged both along the kagome plane and perpendicular to it. However, this structural modulation does not induce a significant anomaly in the acoustic phonons. Our observations suggest an EPC-driven structural instability in CsCr$_3$Sb$_5$ is unlikely and points to an unconventional CDW-like structural modulation.
	
	\begin{acknowledgments}
		The X-ray scattering work was supported by the U.S. Department of Energy, Office of Basic Energy Sciences, under award no. DE-SC0012311 (P.D.), and by Grant-in-Aid for Scientific Research (KAKENHI) (No. JP22H00105), Grant-in-Aid for Scientific Research on innovative areas ``Quantum Liquid Crystals'' (No. JP19H05824) and Grant-in-Aid for Scientific Research for Transformative Research Areas (A) ``Condensed Conjugation'' (No. JP20H05869) from Japan Society for the Promotion of Science (JSPS). The single-crystal growth efforts at Rice are supported by the Robert A. Welch Foundation Grant No. C-1839 (P.D.). The synchrotron X-ray experiment was performed at BL43LXU of the RIKEN SPring-8 Center under the proposal 2024A1108.
	\end{acknowledgments}    
	
	%%Bibliography
	\bibliographystyle{apsrev4-1}
	\bibliography{ccs_ixs_reference}
	
	%%%%%%%%%% Merge with supplemental materials %%%%%%%%%%
	\pagebreak
	\pagebreak
	
	\widetext
	\begin{center}
		\textbf{\large Supplemental Material for ``Absence of Acoustic Phonon Anomaly in a Kagome Metal with Short-ranged Structural Modulation''}
	\end{center}
	
	%%%%%%%%%% Merge with supplemental materials %%%%%%%%%%
	%%%%%%%%%% Prefix a "S" to all equations, figures, tables and reset the counter %%%%%%%%%%
	\setcounter{equation}{0}
	\setcounter{table}{0}
	\setcounter{page}{1}
	\makeatletter
	\renewcommand{\theequation}{S\arabic{equation}}
	\renewcommand{\thefigure}{S\arabic{figure}}
	%\renewcommand{\bibnumfmt}[1]{[S#1]}
	%\renewcommand{\citenumfont}[1]{S#1}
	%%%%%%%%%% Prefix a "S" to all equations, figures, tables and reset the counter %%%%%%%%%%
	\section{CsCr$_3$Sb$_5$ single crystal sample}
	
\begin{figure}[h!]
		\includegraphics[width=7cm]{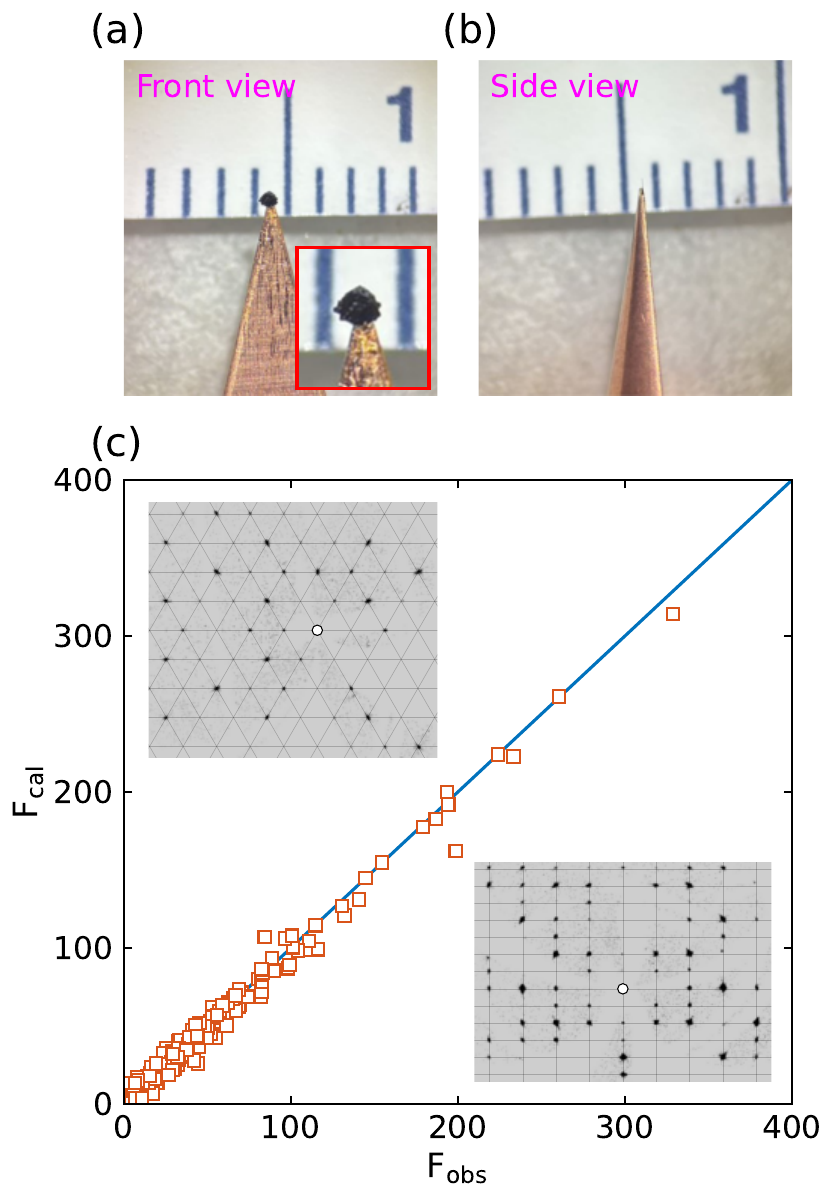}
		\caption{(a) and (b) A piece of CsCr$_3$Sb$_5$ single crystal used in our IXS experiment. [$H$, 0, 0] is approximately along the horizontal direction in (a). [0, 0, $L$] is perpendicular to the large surface. The minimum scale is 1 mm. The inset of (a) is a zoom-in view of the sample. (c) Calculated structure factor F$_{\rm{cal}}$ versus observed structure factor F$_{\rm{obs}}$ from the in-house single crystal XRD measurement at 300 K. Upper left and bottom right insets show the measured XRD patterns of the ($H$, $K$, 0) and (0, $K$, $L$) planes, respectively. The white dots mark the center (0, 0, 0) of the reciprocal space.}
		\label{figs1}
	\end{figure}

	Fig. \ref{figs1}(a) and (b) show one piece of CsCr$_3$Sb$_5$ single crystal used in our inelastic X-ray scattering (IXS) experiment. The sample was attached to the tip of a copper sample holder with GE varnish.% Its ($H$, 0, $L$) plane was put approximately in horizontal with the incident X-ray beam initially close to the [0, 0, $L$] direction.
	One notes that the extreme thinness of the sample (97.5$\%$ transmission) reduced scattering rates by about a factor of 20, or more, compared to what an optimal ($\sim$120 $\mu$m) thickness sample could have provided. Before doing the IXS measurements, the single crystal was checked with an in-house X-ray diffractometer (Rigaku Synergy-S). Crystal structure refinement was performed with the Jana2006 program \cite{Petricek2014}. The obtained space group is $P$6/$mmm$, which is consistent with the previous report \cite{LiuNature2024}. The comparison between the observed and calculated structure factors is presented in Fig. \ref{figs1}(c), where the insets show representative X-ray diffraction (XRD) patterns for the ($H$, $K$, 0) and (0, $K$, $L$) planes. More crystallographic information can be found in Table \ref{tb1} and Table \ref{tb2}. These results confirm the good quality of our CsCr$_3$Sb$_5$ single crystal.
	
	\section{Additional elastic scattering data}
	
\begin{figure}[t!]
		\includegraphics[width=9cm]{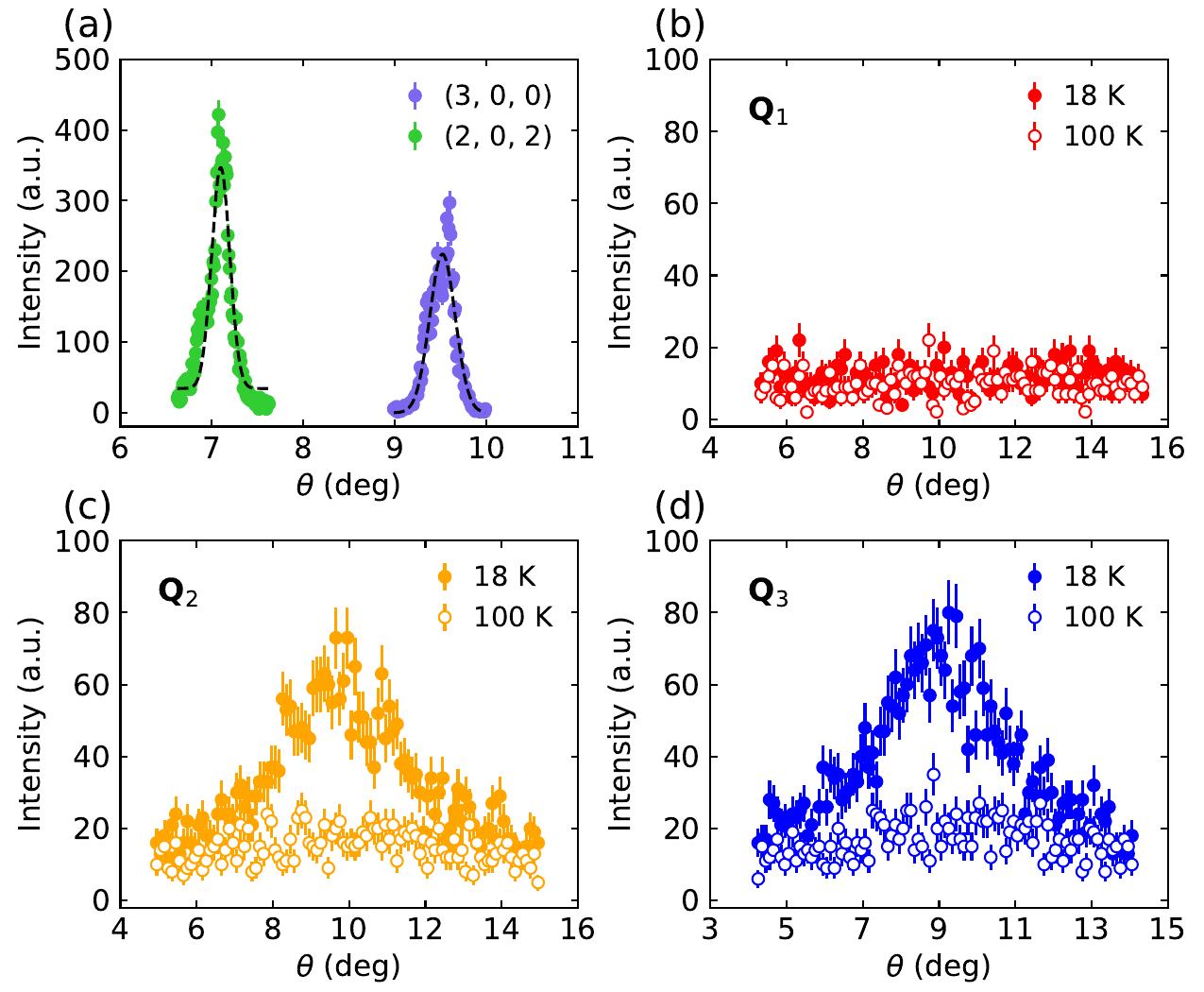}
		\caption{(a) $\theta$-scans performed around (3, 0, 0) and (2, 0, 2) at 18 K, which are total scattered intensity scans. Dashed curves are the fits with Gaussian profiles. (b)-(d) $\theta$-scans performed around $\mathbf{Q}_1$, $\mathbf{Q}_2$, and $\mathbf{Q}_3$, respectively, at 18 K and 100 K. These three scans were done by setting the energy transfer to zero.}
		\label{figs2}
	\end{figure}

	$\theta$-scans performed around (2, 0, 2) and (3, 0, 0) at 18 K are shown in Fig. \ref{figs2}(a). From the full-width-at-half-maximum (FWHM) of their Gaussian fits, we can estimate the overall sample mosaic spread to be  0.33(1)$^{\circ}$ and 0.23(1)$^{\circ}$ for (2, 0, 2) and (3, 0, 0), respectively.
	
	Additional $\theta$-scans performed around $\mathbf{Q}_1$, $\mathbf{Q}_2$, and $\mathbf{Q}_3$ are shown in Fig. \ref{figs2}(b), (c), and (d), respectively. The locations of these momentum transfer positions can be found in Fig. 1(c) of the main text. For $\mathbf{Q}_1$, we can see that there is no peak regardless of 18 K and 100 K. However, for $\mathbf{Q}_2$ and $\mathbf{Q}_3$, broad peaks emerge at 18 K, despite the absence of an evident peak at 100 K. These diffraction features are consistent with the results shown in the main text. We note that due to the short-ranged nature of the structural modulation, the peak widths of $\mathbf{Q}_2$ and $\mathbf{Q}_3$ at 18 K are much larger than those of fundamental structural Bragg peaks [Fig. \ref{figs2}(a)].
	
	\section{Fitting of inelastic X-ray scattering spectra}
	
	At a specific momentum transfer position $\mathbf{Q}$, the experimentally detected energy-dependent IXS intensity $I(E)$ can be approximately decomposed into two parts \cite{MiaoPRX2018,DashwoodPRB2019,LiPRX2021,SongPRB2023}
	\begin{equation}
		I(E)=\sum_{i}S_i(E)*R(E) + aR(E),
	\end{equation}
	where the first term is a sum of the dynamic structure factor $S (E)$ of multiple phonons convoluted with the energy resolution function $R (E)$, and the second term is the elastic scattering contribution represented by the scaled energy resolution function. Since the detector background is generally negligible, a constant background term (e.g., used in \cite{MiaoPRX2018}) is omitted here. For a phonon mode $i$, its dynamic structure factor $S_i(E)$ is related to the imaginary part of the dynamic susceptibility $\chi_i^{\prime\prime}(E)$ through the fluctuation-dissipation theorem
	\begin{equation}
		S_i(E)=\frac{\chi_i^{\prime\prime}(E)}{1-e^{-E/k_{\rm{B}}T}}.
	\end{equation}
	$\chi_i^{\prime\prime}(E)$ can be expressed as the damped harmonic oscillator form
	\begin{equation}
		\chi_i^{\prime\prime}(E) = \frac{C_i\gamma_iEE_i}{(E^2-E_i^2)^2+4E^2\gamma_i^2},
	\end{equation}
	where $C_i$, $E_i$, and $\gamma_i$ are the amplitude, energy, and FWHM of the phonon mode $i$. The energy resolution $R (E)$ is modeled with a pseudo-Voigt function \cite{LeTaconNP2014,MiaoPRX2018,DashwoodPRB2019,LiPRX2021,MiaoNC2023,SongPRB2023,CaoNC2023,ShenNC2023}.
	%\begin{equation}
	%R(E) = (1-\alpha)\frac{A}{\sqrt{2\pi}\sigma}e^{-E^2/(2\sigma^2)}+\alpha \frac{A}{\pi}\frac{\gamma}{E^2+\gamma^2},
	%\end{equation}
	%where $\alpha \in [0, 1]$ determines the relative fraction of the Lorentzian and Gaussian functions, $\sigma$ and $\gamma$ are the widths of the two functions, and $A$ is a scale factor.
	We note that the pseudo-Voigt function is sufficient for the present data and analysis (see also \cite{LeTaconNP2014,MiaoPRX2018,DashwoodPRB2019,LiPRX2021,MiaoNC2023,SongPRB2023,CaoNC2023,ShenNC2023}), but a more careful data treatment could use the deconvolved response as described in \cite{IshikawaJSR2021}.
	
\begin{figure}[b!]
		\includegraphics[width=12cm]{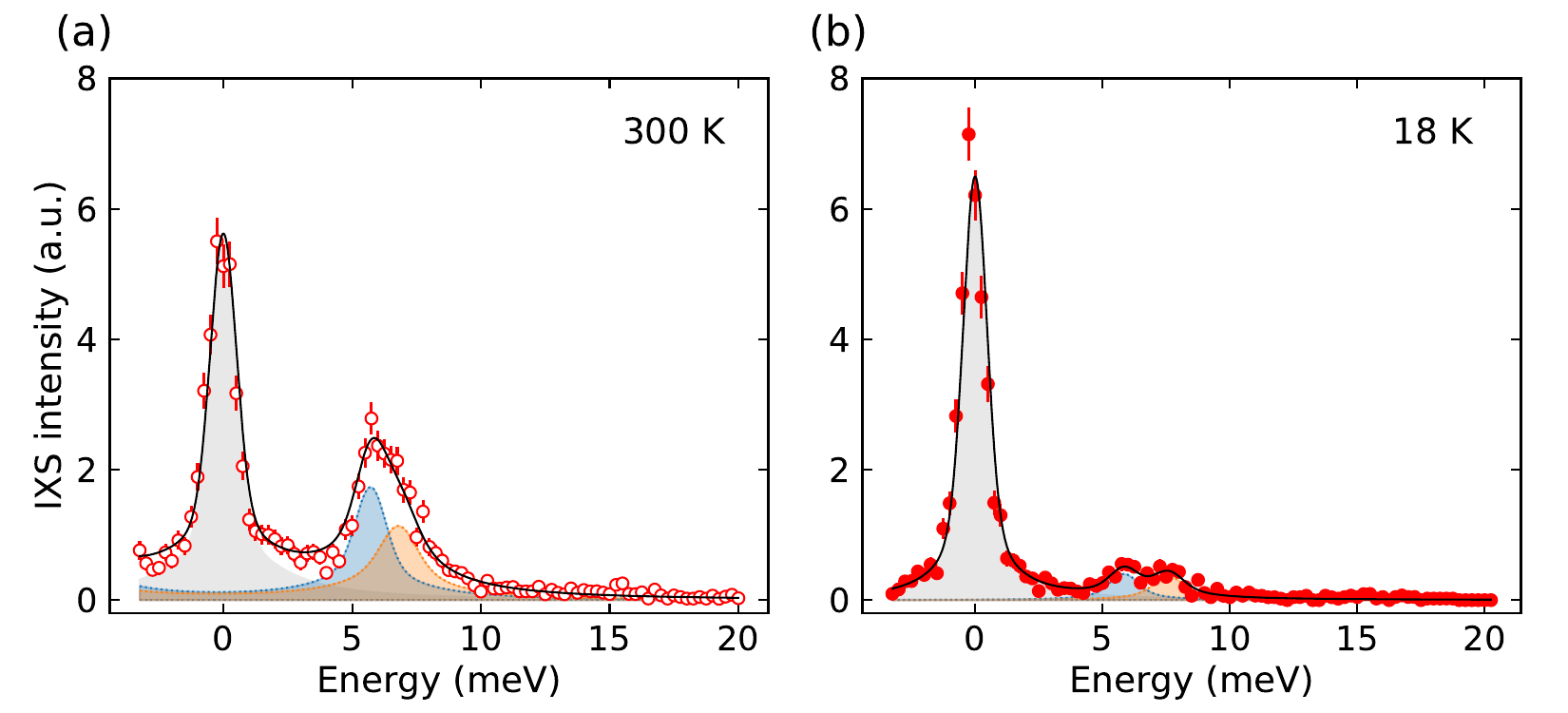}
		\caption{Energy dependence of measured IXS intensities at $\mathbf{Q}_1$ at 300 K (a) and 18 K (b). Solid black curves are the fits as described in the text. The shaded grey areas represent the elastic scattering contributions. The shaded blue and orange areas represent two major phonon modes. Zero intensity is indicated by dashed horizontal lines.}
		\label{figs3}
	\end{figure}

\begin{figure}[b!]
		\includegraphics[width=5cm]{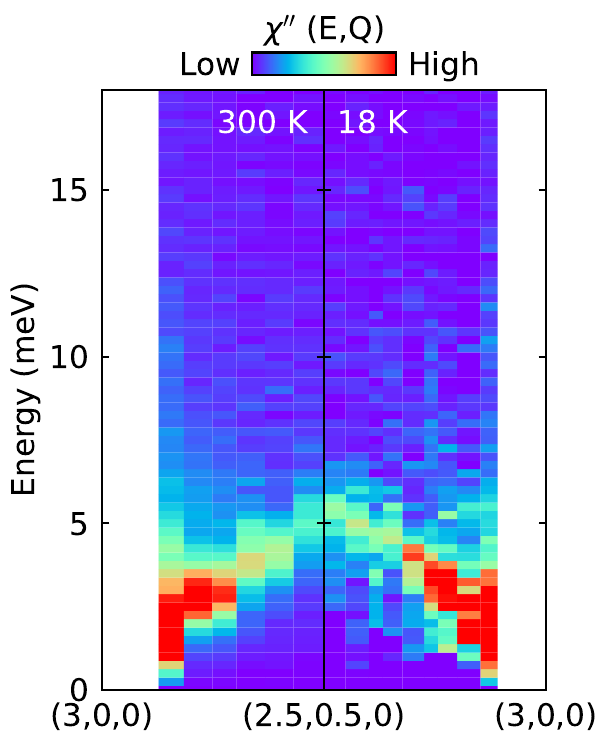}
		\caption{Color maps of the imaginary part of the dynamic susceptibilities at 300 K (left) and 18 K (right) from (3, 0, 0) to (2.5, 0.5, 0).  The intensities are shown in a linear scale.}
		\label{figs4}
	\end{figure}

	We fitted the experimentally measured IXS spectra with equation (1). Fig. \ref{figs3} shows a representative fitting on the IXS spectra of $\mathbf{Q}_1$ at both 300 K and 18 K, where two phonon modes are used. We can see that most major features are captured by our fitting. The obtained overall peak width of $R(E)$ is around 1.2 meV, consistent with the energy resolution. The two phonon peak positions are shown in Fig. 3(c) of the main text. The measured imaginary part of the dynamic susceptibilities presented in the main text was calculated as
	\begin{equation}
		\chi^{\prime\prime}(E)=[I(E) - aR(E) - I_{0}][1-e^{-E/k_{\rm{B}}T}].
	\end{equation}
	Fig. \ref{figs4} shows the color map of the imaginary part of the dynamic susceptibility from (3, 0, 0) to (2.5, 0.5, 0), which goes through $\mathbf{Q}_3$ [see Fig. 1(c) of the main text]. We can confirm that there is no substantial change in the acoustic phonons between 300 K and 18 K.
	
	\section{Estimation of longitudinal sound velocity}
	In the long-wavelength limit (with the wave vector $k \rightarrow 0$), the phonon energy $E$ is approximately linear to $k$ \cite{Kittel2018}
	\begin{equation}
		E=Ck,
	\end{equation}
	where $C$ is a slope constant that can be determined from the experimentally measured phonon dispersion. Here, $E$ is in the unit of meV and $k$ is in the reciprocal lattice unit (r.l.u.) (see the definition of momentum transfer position in the main text). For the three kagome metals CsCr$_3$Sb$_5$, CsV$_3$Sb$_5$, and FeGe, the slope constants $C$s can be obtained from their phonon dispersions close to the Brillouin zone center. To estimate the longitudinal sound velocities, we use the dispersions of their longitudinal acoustic (LA) phonon along $\mathbf{a}^{\ast}$ direction [Fig. 1(c) of the main text]. The corresponding slope constants $C$s in the long-wavelength limit are listed in Table \ref{tb3}. The errors arise from the uncertainties in determining the energy positions of the LA phonon modes. The sound velocity in m$\cdot$s$^{-1}$ is directly related to $C$ through
	\begin{equation}
		v=20.95 \, aC,
	\end{equation}
	where $a$ is the in-plane lattice constant in \AA. The calculated sound velocities for the three kagome metals are listed in Table \ref{tb3}. After knowing the longitudinal sound velocity, the Young’s modulus \cite{Kittel2018} can be further obtained according to 
	\begin{equation}
		E_{\rm{Y}}=\rho v^2,
	\end{equation}
	where $\rho$ is the density. The obtained Young’s moduli are also listed in Table \ref{tb3}, from which we can see CsCr$_3$Sb$_5$ has the smallest Young’s modulus.
	
	\begin{table}[h]
		\caption{
			Crystal data and structure refinement of CsCr$_3$Sb$_5$.
		}
		\begin{ruledtabular}
			\begin{tabular}{cc}
				Empirical formula&CsCr$_3$Sb$_5$\\
				Formula weight&897.64\\
				Temperature&298.23(11) K\\
				Wavelength&0.71073 \AA\\
				Crystal system&Hexagonal\\
				Space group&$P$6/mmm\\
				Unit cell dimensions&a = b = 5.4971(3) \AA, c = 9.2882(7) \AA\\
				&$\alpha$ = $\beta$ = 90$^{\circ}$, $\gamma$ = 120$^{\circ}$\\
				Volume&243.07(3)\AA$^3$\\
				Z&1\\
				Density (calculated)&6.132 g$\cdot$cm$^{-3}$\\
				Absorption coefficient&20.496 mm$^{-1}$\\
				F(000)&382\\
				Crystal size&0.5 $\times$ 0.5 $\times$ 0.01 mm$^3$\\
				Theta range for data collection&4.2809$^{\circ}$ to 36.2766$^{\circ}$\\
				Index ranges&-8 $\le$ $H$ $\le$ 8, -8 $\le$ $K$ $\le$ 6, -15 $\le$ $L$ $\le$ 13\\
				Reflections collected&2157\\
				Independent reflections&280 [R(int) = 0.0810]\\
				Completeness to theta = 26.32$^{\circ}$&99.27 \%\\
				Refinement method&Full-matrix least-squares on F$^2$\\
				Data / restraints / parameters&280 / 0 / 6\\
				Goodness-of-fit on F$^2$&3.78\\
				Final R indices [I $>$ 3sigma(I)]&R1 = 0.0818, wR2 = 0.0893\\
				R indices (all data)&R1 = 0.1037, wR2 = 0.0922\\
				Extinction coefficient&NA\\
			\end{tabular}	
		\end{ruledtabular}
		\label{tb1}
	\end{table}
	
	\begin{table}[h]
		\caption{Atomic coordinates and equivalent isotropic displacement parameters of CsCr$_3$Sb$_5$ at room temperature.}
		\begin{ruledtabular}
			\begin{tabular}{ccccccc}
				\textbf{Atom}&\textbf{Wyckoff.}&\textbf{Occ.}&\textbf{x}&\textbf{y}&\textbf{z}&\textbf{U$\rm_{eq}$}\\
				\midrule
				Cs1&1$a$&1&0&0&0&0.0311(8)\\
				Cr1&3$g$&1&0.5&0.5&0.5&0.0184(10)\\
				Sb1&1$b$&1&0&0&0.5&0.0219(7)\\
				Sb2&4$h$&1&0.3333&0.6667&0.2633(2)&0.0209(5)\\
			\end{tabular}	
		\end{ruledtabular}
		\label{tb2}
	\end{table}
	
	\begin{table}[h]
		\caption{Slope constants of long-wavelength limit LA phonon, in-plane lattice constants, densities, sound velocities, and Young’s moduli for three kagome metals.}
		\begin{ruledtabular}
			\begin{tabular}{cccccc}
				Kagome metal&$C$ (meV$\cdot$r.l.u.$^{-1}$)&$a$ (\AA)&
				$\rho$ (kg$\cdot$m$^{-3}$)&$v$ (m$\cdot$s$^{-1}$)&$E_{\rm{Y}}$ (Pa)\\
				\midrule
				CsCr$_3$Sb$_5$ (this work)&28(1) &5.497 &6.132$\times$10$^{3}$&3.2(1) $\times$ 10$^3$&6.3(4) $\times$ 10$^{10}$\\
				CsV$_3$Sb$_5$&31(2) \cite{XiePRB2022}&5.495 \cite{OrtizPRM2019}&6.102$\times$10$^{3}$ \cite{OrtizPRM2019}&3.6(2) $\times$ 10$^3$&7.9(9) $\times$ 10$^{10}$\\
				FeGe&50(2) \cite{TengNP2023,TengPRL2024}&4.985 \cite{TengNature2022}&7.344$\times$10$^{3}$ \cite{TengNature2022}&5.2(2) $\times$ 10$^3$&19(2) $\times$ 10$^{10}$\\
			\end{tabular}	
		\end{ruledtabular}
		\label{tb3}
	\end{table}

	\clearpage
\end{document}